\documentclass[12pt]{article}
\newcommand{\be}{\begin{equation}}
\newcommand{\ee}{\end{equation}}
\newcommand{\bea}{\begin{eqnarray}}
\newcommand{\eea}{\end{eqnarray}}
\newcommand{\sptwo}{1.4}
\newcommand{\doublespace}{\edef\baselinestretch{\sptwo}\Large\normalsize}
\newcommand{\newsection}[1]{\section{#1}\setcounter{equation}{0}}

\newcounter{newapp}
\begin{document}
\vspace*{0.2in}
\begin{center}
{\large\bf Nonlinear Realization of Supersymmetric AdS Space Isometries}\\
\end{center}
\vspace{0.2in}
\begin{center}
{T.E. Clark}\footnote{e-mail address: clark@physics.purdue.edu}  and {S.T. Love}\footnote{e-mail address: loves@physics.purdue.edu}\\
~\\
{\it Department of Physics\\
Purdue University\\
West Lafayette, IN 47907-1396}\\

\end{center}

\vspace{0.2in}
\begin{abstract}
\noindent
The isometries of $AdS_5$ space and supersymmetric $AdS_5\otimes S_1$ space are nonlinearly realized on  four dimensional Minkowski space. The resultant effective actions in terms of the Nambu-Goldstone modes are constructed. The dilatonic mode governing the motion of the Minkowski space probe brane into the covolume of supersymmetric $AdS_5$ space is found to be unstable and the bulk of the $AdS_5$ space is unable to sustain the brane. No such instablility appears in the non-supersymmetric case.
\end{abstract} 
\hspace*{2in}

\newpage

\doublespace

\newsection{Introduction}

In recent years, there has been a resurgence of interest in conformal field theories fueled by a deeper understanding of properties of supersymmetric (SUSY) gauge theories. This is particularly the case concerning their connection with theories defined in Anti-de Sitter ($AdS$) space\cite{M}-\cite{review}. $AdS_5$ space is defined to be the hyperboloid satisfying the equation
\bea
\frac{1}{m^2} &=& (X^0)^2 -(X^1)^2 -(X^2)^2-(X^3)^2 -(X^4)^2 +(X^5)^2  
\eea
which is embedded in a $6$-dimensional space with invariant interval 
\bea
ds^2 &=& dX^{\cal M} \hat{\eta}_{\cal{M}\cal{N}}dX^{\cal N} ~;~{\cal{M} ,\cal{N}} = 0,1,2,3,4,5 
\eea
\noindent
and  metric tensor $\hat{\eta}_{\cal{M}\cal{N}}$ with signature $(-1,+1,+1,+1,+1,-1)$. 
The isometry group of the hyperboloid is $SO(4,2)$  
whose generators, $\hat{M}^{\cal{M}\cal{N}}=-\hat{M}^{\cal{N}\cal{M}}$ satisfy the algebra:
\bea
[\hat{M}_{\cal{M}\cal{N}}, \hat{M}_{\cal{L}\cal{R}}]&=& i(\hat{\eta}_{\cal{M}\cal{L}}\hat{M}_{\cal{N}\cal{R}}-\hat{\eta}_{\cal{M}\cal{R}}\hat{M}_{\cal{N}\cal{L}}-\hat{\eta}_{\cal{N}\cal{L}}\hat{M}_{\cal{M}\cal{R}}+\hat{\eta}_{\cal{N}\cal{R}}\hat{M}_{\cal{N}\cal{L}})  
\eea

Alternatively  (pseudo-) translation generators can be defined as
\bea
\hat{P}_M &=& m\hat{M}_{5 M}~~;~~ M=0,1,2,3,4  
\eea
so that the $SO(4,2)$ algebra takes the form 
\bea
[\hat{M}_{MN}, \hat{M}_{LR}] &=& i(\eta_{ML}\hat{M}_{NR}-\eta_{MR}\hat{M}_{NL}
-\eta_{NL}\hat{M}_{MR}
+\eta_{NR}\hat{M}_{ML})\cr
[\hat{M}_{MN},\hat{P}_L]&=& i(\eta_{ML}\hat{P}_N
-\eta_{NL}\hat{P}_{M})\cr
[\hat{P}_M, \hat{P}_N] &=& -im^2 \hat{M}_{MN} 
\label{SO42alg}
\eea
where $\eta_{MN}$ is the 5-dimensional Minkowski space metric with \\
signature $(-1, +1, +1, +1, +1)$. Note that 
in the limit $m^2 \rightarrow 0$, this  reduces to the Poincar\'e algebra  of 5-dimensional Minkowski space ($M_5$).

The nonlinear realization of this isometry group which encapsulates the long wavelength dynamical constraints imposed by the spontaneous symmetry breaking when an $AdS_4$ space is embedded in $AdS_5$ space was previously constructed\cite{CLNtV1}. Using coset methods\cite{CCWZ}-\cite{VO}, an $SO(4,2)$  invariant action  in terms of the Nambu-Goldstone modes, $\phi$ and $ v^\mu~,\mu =0,1,2,3~$associated with the spontaneously broken generators $\hat{P}_4$ and $\hat{M}_{\mu 4}$ respectively was secured as 
\bea
S&=&-\sigma \int d^4x ~(det~e) \cr
&=&-\sigma \int d^dx ~(det~\bar{e})~ [cosh(m\phi)]^4 [cos(\sqrt{v^2})+v^\nu\frac{sin(\sqrt{v^2})}{\sqrt{v^2}cosh(m\phi)}{\cal D}_\nu \phi ]  
\eea
where $\sigma$ is the $AdS_4$ brane tension. Here  
\bea
\bar{e}_\nu~^\mu(x) &=& \frac{sinh(\sqrt{m^2x^2})}{\sqrt{m^2 x^2}} P_{\perp~\nu}^\mu (x) +P_{\parallel~\nu}^\mu (x) 
\eea
is the $AdS_4$ vielbein and  ${\cal D}_\mu =\bar{e}_\mu^{-1~\nu}\partial_\nu$ is the $AdS$ covariant derivative while 
\bea
P_{T~\mu\nu}(x)&=&\eta_{\mu\nu}-\frac{x_\mu x_\nu}{x^2} \cr
P_{L~\mu\nu}(x)&=&\frac{x_\mu x_\nu}{x^2}  
\eea
are transverse  and longitudinal projectors respectively. 

Since this action is independent of $\partial_\mu v$, $v^\mu$ can be eliminated via its field equation  
\bea
&&v^\nu\frac{tan(\sqrt{v^2})}{\sqrt{v^2}}= \eta^{\mu\nu}\frac{{\cal D}_\nu \phi}{cosh(m\phi)}  
\eea
 and   the $SO(4,2)$ invariant action can be recast as 
\be
S= -\sigma \int d^4x ~(det~\bar{e})~ [cosh(m\phi)]^4 \sqrt{1+\frac{{\cal D}_\mu\phi \eta^{\mu\nu}{\cal D}_\nu\phi}{cosh^2(m\phi)}} ~.
\ee
Note that  the Nambu-Goldstone mode, $\phi$, contains a mass term with $m^2_\phi =  4m^2 $ as well as  non-derivative interactions.
The action constitutes an $AdS$ generalization of the Nambu-Goto action: \
\bea
S_{NG}&=& -\sigma\int d^4x \sqrt{1+(\partial_\mu \phi)^2} ~.
\eea

Using the factorized form of the $AdS_{5}$ vielbein, along with $v^\mu$ field equation, the invariant interval for $AdS_{5}$  space reads
\bea
ds^2 &=&  e^{2A(\phi)}dx^\mu dx^\nu \eta^{\rho\mu} \bar{e}_{\nu\rho}(x) +(d\phi(x))^2 
\eea
with warp factor $A(\phi)=\ell n[cosh(m\phi)]$. This allows the identification of the Nambu-Goldstone mode $\phi$ with the covolume coordinate describing the motion of the $AdS_4$ brane into the remainder of the $AdS_{5}$ space. 

In this paper, we construct the nonlinear realization of the $AdS_5$ and SUSY $AdS_5 \otimes S_1$ isometry groups on an embedded  four dimensional Minkowski space. The supersymmetric case turns out to be particulary interesting. Here it is found that the Nambu-Goldstone boson describing the motion of the inserted  Minkowski space probe brane into the remainder  of the $AdS_5$ space exhibits an instability which drives the probe brane to $-\infty$. Alternatively, this result can be interpreted as the incompatibilty of simultaneous nonlinear realizations of both scale symmetry and supersymmetry or the nonviability of the spectrum containing both the dilaton of spontaneously broken scale symmetry  and the Goldstino of spontaneously broken supersymmetry. On the other hand,  no such unstable behavior arises when the $M_4$ probe brane is inserted into non-supersymmetric $AdS_5$ space. 

\newsection{Four dimensional  Minkowski space  probe brane in $AdS_{5}$ space}

To study the case of the 4-dimensional Minkowski space probe brane in $AdS_5$ space, it proves useful to introduce the 
$AdS_{5}$ coordinates:  
\bea
X^\mu &=& e^{mx_4}x^\mu \cr
X^4&=& \frac{1}{m}[sinh(mx_4)-\frac{m^2x^2}{2}e^{mx_4}]\cr
X^{5}&=&\frac{1}{m}[cosh(mx_4)+\frac{m^2x^2}{2}e^{mx_4}]
\eea
which  parametrize the hyperboloid and in terms of which the $AdS_5$ space 
 invariant interval takes the form: 
\be
ds^2= e^{2mx_4}dx^\mu \eta_{\mu\nu}dx^\nu +(dx_4)^2 ~.
\ee

Inserting the  Minkowski space probe brane at $x_4 =0$, the 
broken generators are identified as  $\hat{P}^4 \equiv mD ~,~ \hat{M}^{\mu 4}\equiv m{\cal K}^\mu$. Defining 
\bea
P^\mu &=&  \hat{P}^\mu +m\hat{M}^{\mu 4}\cr
M^{\mu\nu}&=&\hat{M}^{\mu\nu}
\eea 
the $SO(4,2)$ algebra, Eq. (\ref{SO42alg}), takes the form
\bea
&&[P^\mu, P^\nu]=0~;~[M^{\mu\nu}, P^\lambda]=i(\eta^{\mu\lambda}P^\nu-\eta^{\nu\lambda}P^\mu)\cr
&&[M^{\mu\nu}, M^{\lambda\rho}]=i(\eta^{\mu\lambda}M^{\nu\rho}-\eta^{\mu\rho}M^{\nu\lambda}-\eta^{\nu\lambda}M^{\mu\rho}+\eta^{\nu\rho}M^{\mu\lambda})\cr
&&[{\cal K}^\mu, {\cal K}^\nu] =\frac{i}{m^2}M^{\mu\nu}~;~[M^{\mu\nu}, {\cal K}^\lambda]=i(\eta^{\mu\lambda}{\cal K}^\nu-\eta^{\nu\lambda}{\cal K}^\mu) \cr
&&[P^\mu, {\cal K}^\nu]= -i(\eta^{\mu\nu}D -M^{\mu\nu})\cr
&&[D, P^\mu]=iP^\mu ~;~[D, {\cal K}^\mu]=\frac{i}{m^2}(P^\mu-m^2{\cal K}^\mu) ~;~[D, M^{\mu\nu}]= 0 
\label{msa}~.
\eea
Note that $M^{\mu\nu}$ and $P^\mu$ form a  Poincar\'e algebra, while the generators ${\cal K}^\mu, M^{\mu\nu}$ constitute an $SO(3,2)$  subalgebra\cite{Ferrara}-\cite{Ivanov03}. 

A model independent way of encapsulating the long wavelength dynamical constraints imposed by spontaneous symmetry breakdown is to realize this $SO(4,2)$ isometry nonlinearly  on the Nambu-Goldstone bosons consisting of the dilaton, $\varphi$, associated with the broken symmetry generator $D$ and $v^\mu$ associated with the ${\cal K^\mu}$ spontaneously broken generators.  Since the spontaneously broken symmetries are space-time symmetries, the motion in the coset space is accompanied by  a motion in space-time. Thus we consider the product of a space-time translation group element with the coset group element and define the group element:
\bea
&& \Omega = e^{-ix^\mu P_\mu}e^{i\varphi D} e^{-iv^\mu {\cal K}_\mu} ~. 
\eea
To extract the total variations of the coset coordinates and the corresponding transformation of the space-time point, 
consider the product $ g\Omega $
with $g$ a general group element parametrized by real infinitesimal constants. An explicit calculation then gives 
\bea
 g \Omega(x, \phi(x), v(x)) &=&\Omega (x^\prime, \varphi^\prime(x^\prime), v^\prime(x^\prime)) h(\theta) 
\label{gt}
\eea
 with $h=e^{\frac{i}{2}\theta^{\mu\nu}(x){M}_{\mu\nu}}$  an element of the unbroken (stability) group. 
 So doing, allows the extraction of the forms of $x^\prime~,~ \varphi^\prime(x^\prime)~,~v^{\mu~\prime}(x^\prime)~,~\theta^{\mu\nu}(x)$\cite{Ivanov02}-\cite{Ivanov03}.\\

To construct $SO(4,2)$ invariants, it is useful to define the algebra valued Maurer-Cartan 1-form $\Omega^{-1}d\Omega$ which, using Eq. (\ref{gt}), is seen to have the  simple transformation property
\bea
(\Omega^{-1}d\Omega)^\prime (x^\prime)&=&[h(\Omega^{-1}d\Omega) h^{-1}](x)+(h dh^{-1})(x) 
\label{MCTL}
\eea
Expanding the Maurer-Cartan form in terms of the generators as 
\bea
\Omega^{-1}d\Omega (x) &=& i[-\omega^\mu_P(x) P_\mu +\omega_D(x) D -\omega^\mu_{\cal K}(x) {\cal K}_\mu+\frac{1}{2}\omega^{\mu\nu}_M(x) M_{\mu\nu}  ] 
\eea
and exploiting the $SO(4,2)$ algebra along with liberal application of the Baker-Campbell-Haussdorff formula,  the various 1-form coefficients \\
$\omega^\mu_P(x)=dx^\nu e_\nu^{~\mu}~,~\omega_D(x)~,~\omega^\mu_{\cal K}(x)~,~\omega_M^{\mu\nu}$  are secured. Here 
\bea
e_\mu^{~\nu}&=& e^\varphi [P_{\perp\mu}^\nu(v) +P_{\parallel\mu}^\nu(v)cos(\sqrt{m^2/v^2})]-\partial_\mu\varphi v^\nu \frac{sin(\sqrt{m^2/v^2})}{\sqrt{m^2/v^2}} 
\eea
is the $AdS_5$ vielbein.

Again using the $SO(4,2)$ algebra, this time in the above transformation law (\ref{MCTL}), leads to the invariant combination
\bea 
d^4x^\prime ~det ~e^\prime = d^4x ~det~e  ~.
\eea
Thus an $SO(4,2)$ invariant action is constructed  as
\bea
S&=&-\sigma \int d^4x ~det ~e \cr
&=& -\sigma \int d^4x e^{4\varphi }[cos(\sqrt{m^2 v^2})-e^{-\varphi}\partial_\mu \varphi v^\mu \frac{sin(\sqrt{m^2/v^2})}{\sqrt{m^2/v^2}}]  
\eea
with $\sigma$ the Minkowski probe brane tension. As previously, the $v^\mu$ Nambu-Goldstone field is not an independent dynamical degree of freedom and it can be eliminated using its field equation  
\bea
v^\mu \frac{tan(\sqrt{m^2/v^2})}{\sqrt{m^2/v^2}}&=&- e^{-\varphi}\eta^{\mu\nu}\partial_\nu \varphi ~.
\eea
Substituting back then produces the action
\bea
S&=& -\sigma \int d^4x e^{4\varphi } \sqrt{1+\frac{1}{m^2}e^{-2\varphi}\partial_\mu\varphi\eta^{\mu\nu}\partial_\nu\varphi}  ~.
\label{mba}
\eea
After using the $v^\mu$ field equation, the invariant interval 
\bea
ds^2 &=&  dx^\mu e_\mu^{~\lambda}\eta_{\lambda\rho}e_\nu^{~\rho} dx^\nu \cr
&=&e^{2\varphi} dx^\mu\eta_{\mu\nu}dx^\nu +\frac{1}{m^2}(d\varphi)^2  
\eea
is seen to have the  same form as the $AdS_{5}$ invariant interval obtained previously after identification $\varphi\Leftrightarrow\frac{1}{m}x_4$. Thus the dilaton dynamics describes motion of the brane into the rest of $AdS_{5}$ space 

In the above construction, we have chosen a particular
combination of the broken generators, referred to as the maximal solvable subgroup basis or parametrization\cite{Ferrara}, whose nonlinear realization on the Nambu-Goldstone modes has the attractive feature of directly  relating the Nambu-Goldstone dilaton to the motion of the brane into the rest of $AdS_5$ space. An 
alternate choice of broken generators is given by 
\bea
K^\mu &=& \frac{1}{m^2}\hat{P}^\mu -\frac{1}{m}\hat{M}^{\mu 4}=\frac{1}{m^2}(P^\mu -2m^2{\cal K}^\mu) ~.
\eea
This choice leads to the 4-dimensional conformal algebra
\bea
&&[P^\mu, P^\nu]=0~;~[M^{\mu\nu}, P^\lambda]=i(\eta^{\mu\lambda}P^\nu-\eta^{\nu\lambda}P^\mu)\cr
&&[M^{\mu\nu}, M^{\lambda\rho}]=i(\eta^{\mu\lambda}M^{\nu\rho}-\eta^{\mu\rho}M^{\nu\lambda}-\eta^{\nu\lambda}M^{\mu\rho}+\eta^{\nu\rho}M^{\mu\lambda})\cr
&&[K^\mu, K^\nu] =0~;~[M^{\mu\nu}, K^\lambda]=i(\eta^{\mu\lambda}K^\nu-\eta^{\nu\lambda}K^\mu)\cr
&&[P^\mu, K^\nu]= 2i(\eta^{\mu\nu}D -M^{\mu\nu})\cr
&&[D, P^\mu]=iP^\mu ~;~[D, K^\mu]=-iK^\mu ~;~[D, M^{\mu\nu}]= 0 ~.
\eea
Since the generators $K^\mu$ and ${\cal K}^\mu$ differ only by the unbroken translation generator $P^\mu$, it follows that the action (\ref{mba}) is also invariant under 4-dimensional conformal transformations. Moreover, one can  subtract the invariant action piece $\sigma\int d^dx e^{\varphi d} $ ensuring a zero vacuum energy and thus producing the conformally invariant action
\bea
S&=& -\sigma \int d^dx e^{\varphi d} [\sqrt{1+\frac{1}{m^2}e^{-2\varphi}\partial_\mu\eta^{\mu\nu}\varphi\partial_\mu\varphi}-1]  ~.
\label{mba1}
\eea
Note that the leading term in a momentuum expansion is simply
\bea
S&=& -\sigma \int d^dx e^{2\varphi} \partial_\mu\varphi\eta^{\mu\nu}\partial_\mu\varphi 
\eea
which is the familiar result.

\newsection{Four dimensional Minkowski space probe brane in SUSY $AdS_5\otimes S_1$ space}

The supersymmetric $AdS_5 \otimes S_1$ isometry algebra includes the generators $\hat{M}^{MN}~,~\hat{P}^M;$\\
$M,N=0,1,2,3,$ of the $SO(4,2)$ isometry algebra, the 
SUSY fermionic charges ${\cal Q}_a, \bar{\cal{Q}}_b~;~a,b =1,2,3,4$ and the 
 $R$ charge which is the  generator of the $U(1)$ isometry of $S_1$. This $SU(2,2|1)$ isometry algegra\cite{Zumino}-\cite{vanHolten} is 
\bea
&&[\hat{M}^{MN}, \hat{M}^{LR}]= i(\eta^{ML}\hat{M}^{NR}-\eta^{MR}\hat{M}^{NL}-\eta^{NL}\hat{M}^{MR}+\eta^{NR}\hat{M}^{ML}) \cr
&&[\hat{M}^{MN}, \hat{P}^L]=i(\eta^{ML}\hat{P}^N -\eta^{NL}\hat{P}^M) \cr
&&[\hat{P}^M, \hat{P}^N]=-im^2 \hat{M}^{MN} \cr
&&[\hat{M}^{MN}, {\cal Q}_a]=-\frac{1}{2}(\Sigma^{MN}{\cal Q})_a ~;~[\hat{M}^{MN}, \bar{\cal Q}_a]=\frac{1}{2}(\bar{\cal Q}\Sigma^{MN})_a \cr
&&[\hat{P}^M, {\cal Q}_a]= -\frac{m}{2}(\Gamma^\mu {\cal Q})_a ~;~[\hat{P}^M, \bar{\cal Q}_a]= \frac{m}{2}(\bar{\cal Q}\Gamma^M )_a \cr
&&[R, {\cal Q}_a] ={\cal Q}_a ~~;~~[R, \bar{\cal Q}_a] =-\bar{\cal Q}_a \cr
&&\{{\cal Q}_a, \bar{\cal Q}_b \}= 2(\Gamma^M_{ab} \hat{P}_M -\frac{m}{2}\Sigma^{MN}_{ab} \hat{M}_{MN}-\frac{3}{2}m \delta_{ab}R) 
\eea
where the five dimensional  $4\times4$ matrices  $\Gamma^M$ satisfy the Clifford algebra
\bea
&&\{\Gamma^M, \Gamma^N\}=-2\eta^{MN} 
\eea
and the spin matrices are
\bea
&&\Sigma^{MN}=\frac{i}{2}[\Gamma^M, \Gamma^N]=-\Sigma^{NM} ~.
\eea
We choose
\bea
\Gamma^M = \biggr\{ \begin{array}{clcr}
\gamma^\mu&~;~& M=\mu =0,1,2,3\cr
i\gamma_5 &~;~& M=4 \end{array} 
\eea
and use a Weyl representation for the $\gamma$ matrices so that
\bea
\gamma^\mu=\biggr(\begin{array}{clcr}
0 && \sigma^\mu_{\alpha \dot\alpha}\cr
\bar\sigma^{\mu~\dot\alpha\alpha} &&0 \end{array}\biggr)~~;~~\alpha, \dot\alpha=1,2~~;~~\gamma^4=\biggr(\begin{array}{clcr}
-i && 0\\
0 &&i \end{array}\biggr)
\eea
\bea
\Sigma^{\mu\nu}=\biggr(\begin{array}{clcr}
\sigma^{\mu\nu~\beta}_\alpha && 0\cr
0 &&\bar\sigma^{\mu\nu~\dot\alpha}_{~~~~~\dot\beta} \end{array}\biggr)~~;~~\Sigma^{\mu 4}=\biggr(\begin{array}{clcr}
0 && -\sigma^\mu_{\alpha \dot\alpha}\cr
\bar\sigma^{\mu~\dot\alpha\alpha }&&0 \end{array}\biggr)=-\Sigma^{4\mu} ~.
\eea

Embedding a four dimensional Minkowski space ($M_4$) probe brane at $x^4 =0$ breaks the space-time symmetries generated by $\hat{P}^4\equiv mD$ and
$\hat{M}^{\mu4}\equiv  m{\cal K}^\mu$, as well as the supersymmetries generated by 
${\cal Q}_a,~~;~~ \bar{\cal Q}_a $ and the $R$ symmetry. 
Defining $P^\mu =  \hat{P}^\mu +m\hat{M}^{\mu 4}$ and 
\bea
{\cal Q}_a = \biggr( \begin{array}{cc}
 Q_\alpha\cr
-im\bar{S}^{\dot\alpha} \end{array}\biggr )~~;~~
\bar{\cal Q}_a = \biggr( \begin{array}{rr}
imS^\alpha, 
 \bar{Q}_{\dot\alpha} \end{array}\biggr ) 
\eea
the resultant algebra is the SUSY extension of the $P^\mu, M^{\mu\nu}, D, {\cal K}^\mu$ algebra given in Eq. (\ref{msa}) and reads 
\bea
&&[P^\mu, P^\nu]=0~;~[M^{\mu\nu}, P^\lambda]=i(\eta^{\mu\lambda}P^\nu-\eta^{\nu\lambda}P^\mu)\cr
&&[M^{\mu\nu}, M^{\lambda\rho}]=i(\eta^{\mu\lambda}M^{\nu\rho}-\eta^{\mu\rho}M^{\nu\lambda}-\eta^{\nu\lambda}M^{\mu\rho}+\eta^{\nu\rho}M^{\mu\lambda})\cr
&&[{\cal K}^\mu, {\cal K}^\nu] =\frac{i}{m^2}M^{\mu\nu}~;~[M^{\mu\nu}, {\cal K}^\lambda]=i(\eta^{\mu\lambda}{\cal K}^\nu-\eta^{\nu\lambda}{\cal K}^\mu) \cr
&&[P^\mu, {\cal K}^\nu]= -i(\eta^{\mu\nu}D -M^{\mu\nu})~;~[D, P^\mu]=iP^\mu \cr
&&[D, {\cal K}^\mu]=\frac{i}{m^2}(P^\mu-m^2{\cal K}^\mu) ~;~[D, M^{\mu\nu}]= 0\cr
&&[R,P^\mu]= 0~;~[R, {\cal K}^\mu]=0~;~[R, M^{\mu\nu}]= 0 \cr
&&\{Q_\alpha, Q_\beta\} = 0 ~;~\{\bar{Q}_{\dot\alpha}, \bar{Q}_{\dot\beta}\} =0 \cr
&&\{S_\alpha, S_\beta \}=0 ~;~\{\bar{S}_{\dot\alpha}, \bar{S}_{\dot\beta}\} = 0\cr
&&\{Q_\alpha, \bar{Q}_{\dot\alpha}\}= 2\sigma^\mu_{\alpha\dot\alpha}P_\mu~;~\{S_\alpha,\bar{S}_{\dot\alpha}\} = \frac{2}{m^2}\sigma^\mu_{\alpha\dot\alpha}(P_\mu-2m^2 {\cal K}_\mu)\cr
&&\{Q_\alpha,S_\beta \}= i (\sigma^{\mu\nu}_{\alpha\beta}M_{\mu\nu}+2i\epsilon_{\alpha\beta}D+3\epsilon_{\alpha\beta}R)\cr
&&\{\bar{Q}_{\dot\alpha}, \bar{S}_{\dot\beta} \}= -i (\bar\sigma^{\mu\nu}_{\dot\alpha\dot\beta}M_{\mu\nu}-2i\epsilon_{\dot\alpha\dot\beta}D + 3\epsilon_{\dot\alpha\dot\beta}R)\cr
&&\{Q_\alpha, \bar{S}_{\dot\alpha}\}= 0 ~;~\{S_\alpha, \bar{Q}_{\dot\alpha}\}= 0\cr
&&[P^\mu, Q_\alpha ]= 0 ~;~[P^\mu, \bar{Q}_{\dot\alpha}] = 0\cr
&&[P^\mu, S_\alpha ]= i\sigma^\mu_{\alpha\dot\alpha}\bar{Q}^{\dot\alpha}~;~[P^\mu, \bar{S}_{\dot\alpha}]= iQ^\alpha \sigma^\mu_{\alpha\dot\alpha} \cr
&&[M^{\mu\nu}, Q_\alpha]= -\frac{1}{2}\sigma^{\mu\nu}_\alpha~^\beta Q_\beta ~;~[M^{\mu\nu}, \bar{Q}_{\dot\alpha}]= -\frac{1}{2}\bar{\sigma}^{\mu\nu}_{\dot\alpha \dot\beta} \bar{Q}^{\dot\beta } \cr
&&[M^{\mu\nu}, S_\alpha]= -\frac{1}{2}\sigma^{\mu\nu}~_\alpha~^\beta S_\beta ~;~[M^{\mu\nu}, \bar{S}_{\dot\alpha}]= -\frac{1}{2}\bar{\sigma}^{\mu\nu}~_{\dot\alpha \dot\beta} \bar{S}^{\dot\beta } \cr
&&[R, Q_\alpha]= Q_\alpha ~;~[R, \bar{Q}_{\dot\alpha}] = -\bar{Q}_{\dot\alpha}~;~ \cr
&&[R,S_\alpha ] = -S_\alpha~;~[R,\bar{S}_{\dot\alpha}] = \bar{S}_{\dot\alpha}\cr
&&[D, Q_\alpha]= \frac{i}{2}Q_\alpha ~;~[D, \bar{Q}_{\dot\alpha}]= \frac{i}{2}\bar{Q}_{\dot\alpha}\cr
&&[D,S_\alpha]= -\frac{i}{2}S_\alpha ~;~[D,\bar{S}_{\dot\alpha}]=-\frac{i}{2}\bar{S}_{\dot\alpha} \cr
&&[{\cal K}^\mu, Q_\alpha]= -\frac{i}{2}\sigma^\mu_{\alpha\dot\alpha} \bar{S}^{\dot\alpha} ~;~[{\cal K}^\mu, \bar{Q}_{\dot\alpha}]= -\frac{i}{2}S^\alpha \sigma^\mu_{\alpha\dot\alpha} \cr
&& \cr
&&[{\cal K}^\mu, S_\alpha]=\frac{i}{2m^2}\sigma^\mu_{\alpha\dot\alpha}\bar{Q}^{\dot\alpha} ~;~[{\cal K}^\mu,\bar{S}_{\dot\alpha}]=\frac{i}{2m^2}Q^\alpha\sigma^\mu_{\alpha\dot\alpha} ~.
\eea

Using coset methods, we nonlinearly realize this $SU(2,2|1)$ isometry algebra of the super-$AdS_5 \otimes S_1$ space on the  Nambu-Goldstone modes of the broken symmetries. These are the dilaton, $\varphi$, and $v^\mu$ associated with $D$ and ${\cal K}^\mu$ respectively, the Goldstinos $ \lambda_\alpha~~,~~ \bar{\lambda}_{\dot\alpha}~~$ and $\lambda_{S\alpha}~,~\bar\lambda_{S\dot\alpha}$ of the spontaneously broken supersymmetries, $Q_\alpha, \bar{Q}_{\dot\alpha}, S_\alpha, \bar{S}_{\dot\alpha},$ and the $R$ axion $a$. Note that all the supersymmetries have been broken and there is no residual unbroken SUSY. The Nambu-Goldstone modes, $v^\mu$, associated with the broken symmetries generated by ${\cal K}^\mu$ and the Goldstinos $\lambda_{S\alpha}~;~\bar\lambda_{S\dot\alpha}$ associated with the supersymmetries $S_\alpha, \bar{S}_{\dot\alpha}$ are not independent dynamical degrees of freedom but are instead given in terms of the dilaton, $\varphi$, the Goldstinos, $ \lambda_\alpha~~,~~ \bar{\lambda}_{\dot\alpha}~~$ and the $R$ axion, $a$. Nonetheless, it is still necessary to include them as auxiliary fields in the coset construction of the $SU(2,2|1)$ invariant action. To implement this construction, we define the product of a space-time translation and coset group elements by
\be
\Omega = e^{-ix^\mu P_\mu} e^{i\left[ \lambda^\alpha (x) Q_\alpha +\bar\lambda_{\dot\alpha}\bar{Q}^{\dot\alpha}\right]}e^{i\varphi D}e^{iaR}  e^{i\left[ \lambda_S^\alpha (x) S_\alpha +\bar\lambda_{\bar{S}\dot\alpha}\bar{S}^{\dot\alpha}\right]} e^{-iv^\mu {\cal K}_\mu} .
\ee

The covariant building blocks out of which an invariant action can be constructed are secured using the Maurer-Cartan one-form $\Omega^{-1} d \Omega$, where $d=dx^\mu \partial_\mu$.  Expanding in terms of the $SU(2,2|1)$ charges via
\bea
\Omega^{-1} d \Omega &=& i\left[ -\omega^\mu P_\mu + \omega_Q^\alpha Q_\alpha + \bar{\omega}_{\bar{Q}\dot\alpha} \bar{Q}^{\dot\alpha} + \omega_D D + \omega_R R \right. \cr
 & &\left. \qquad\qquad- \omega_{\cal K}^\mu {\cal K}_\mu + \omega_S^\alpha S_\alpha + \bar{\omega}_{\bar{S}\dot\alpha} \bar{S}^{\dot\alpha} +\frac{1}{2} \omega_M^{\mu\nu} M_{\mu\nu}\right] ,
\eea
the individual covariant one-forms are then extracted as
\bea
\omega^\mu &=& \tilde\omega^\mu +P^\mu_{\parallel \nu} (v) \tilde\omega^\nu \left[ \cos({\sqrt{v^2 /m^2}})-1\right] -\tilde\omega_D \frac{v^\mu}{m^2} \frac{\sin({\sqrt{v^2 /m^2}})}{\sqrt{v^2 /m^2}} \cr
 & &  \cr
\omega_D &=& \tilde\omega_D \cos({\sqrt{v^2 /m^2}})- v_\nu \tilde\omega^\nu \frac{\sin({\sqrt{v^2 /m^2}})}{\sqrt{v^2 /m^2}} \cr
 & & \cr
\omega_R &=& \tilde\omega_R  \cr
 & & \cr
\omega_Q^\alpha &=&  \tilde\omega_Q^\alpha \cos{(\frac{\sqrt{v^2 /m^2}}{2}})+\left(\bar{\tilde\omega}_{\bar{S}}\bar{\sigma}\cdot{v}\right)^{\alpha}
\frac{\sin({\frac{\sqrt{v^2 /m^2}}{2}})}{m^2 \sqrt{v^2 /m^2}}\cr
 & & \cr
\bar\omega_{\bar{Q}\dot\alpha} &=& \bar{\tilde\omega}_{\bar{Q}\dot\alpha} \cos(\frac{\sqrt{v^2 /m^2}}{2}) -\left( \tilde\omega_S \sigma\cdot{v} \right)_{\dot\alpha} \frac{\sin({\frac{\sqrt{v^2 /m^2}}{2}})}{m^2 \sqrt{v^2 /m^2}}  \cr
 & & \cr
\omega_S^\alpha &=& \tilde\omega_S^\alpha \cos({\frac{\sqrt{v^2 /m^2}}{2}})-\left(\bar{\tilde\omega}_{\bar{Q}}\bar{\sigma}\cdot{v}\right)^{\alpha}
\frac{\sin({\frac{\sqrt{v^2 /m^2}}{2}})}{\sqrt{v^2 /m^2}} \cr
 & & \cr
\omega_{\bar{S}\dot\alpha} &=& \bar{\tilde\omega}_{\bar{S}\dot\alpha} \cos({\frac{\sqrt{v^2 /m^2}}{2}})+\left( \tilde\omega_Q \sigma\cdot{v} \right)_{\dot\alpha} \frac{\sin({\frac{\sqrt{v^2 /m^2}}{2}})}{\sqrt{v^2 /m^2}}\cr
 & & \cr
\omega_{\cal K}^\mu &=& dv^\mu +\tilde\omega_{\cal K}^\mu + \left[ \cos({\sqrt{v^2 /m^2}})-1\right]P_{\perp ~\nu}^\mu (v)\tilde\omega_{\cal K}^\nu 
\cr
 & & \cr
 & &\qquad\qquad+ \left[ \frac{\sin({\sqrt{v^2 /m^2}})}{\sqrt{v^2 /m^2}} -1 \right] P_{\perp \nu}^\mu (v) dv^\nu   \cr
 & & \cr
 & &+m^2 \left[ \cos({\sqrt{v^2 /m^2}})-1\right] \left[ P_{\perp \nu}^\mu (v) \tilde\omega^\nu - P_{\parallel \nu}^\mu (v) \tilde\omega^\nu \right]\cr
 & & \cr
 & &\qquad\qquad\qquad+ \tilde\omega_D v^\mu \frac{\sin({\sqrt{v^2 /m^2}})}{\sqrt{v^2 /^2}}  -\tilde\omega_M^{\mu \nu} v_\nu \frac{\sin({\sqrt{v^2 /m^2}})}{\sqrt{v^2 /m^2}} \cr
 & & \cr
\omega_M^{\mu\nu} &=& \tilde\omega_M^{\mu\nu}-\left[ \cos({\sqrt{v^2 /m^2}})-1\right] \left( \frac{v^\mu dv^\nu -v^\nu dv^\mu}{v^2} \right)  \cr
 & & \cr
 & & \qquad\qquad\qquad-\frac{1}{2} \frac{\sin({\sqrt{v^2 /m^2}})}{\sqrt{v^2 /m^2}} \left( \tilde\omega^\mu v^\nu - \tilde\omega^\nu v^\mu \right)\cr
 & & \cr 
 & & +\frac{1}{2}\left[ \cos({\sqrt{v^2 /m^2}})-1\right] \left( \tilde\omega_M^{\mu\rho} P_{\parallel \rho}^\nu (v) -\tilde\omega_M^{\nu\rho} P_{\parallel \rho}^\mu (v) \right) \cr
 & &\qquad\qquad\qquad -\left( \tilde\omega_{\cal K}^\mu v^\nu -\tilde\omega_{\cal K}^\nu v^\mu\right) \frac{\sin({\sqrt{v^2 /m^2}})}{m^2 \sqrt{v^2 /m^2}}.\cr
 & & 
\label{omega}
\eea
Here the one-forms denoted by the $\tilde\omega$ are defined via the expansion
\bea
 & & \left(e^{-i\left[ \lambda_S^\alpha (x) S_\alpha +\bar\lambda_{\bar{S}\dot\alpha}\bar{S}^{\dot\alpha}\right]} e^{-iaR}e^{-i\varphi D}e^{-i\left[ \lambda^\alpha (x) Q_\alpha +\bar\lambda_{\dot\alpha}\bar{Q}^{\dot\alpha}\right]}e^{ix^\mu P_\mu}\right)\qquad\qquad\qquad\qquad\qquad\cr
 & & \qquad\qquad\qquad\qquad\qquad d ~\left(e^{-ix^\mu P_\mu} e^{i\left[ \lambda^\alpha (x) Q_\alpha +\bar\lambda_{\dot\alpha}\bar{Q}^{\dot\alpha}\right]}e^{i\varphi D}e^{iaR} e^{i\left[ \lambda_S^\alpha (x) S_\alpha +\bar\lambda_{\bar{S}\dot\alpha}\bar{S}^{\dot\alpha}\right]} \right)\cr
 &=& i\left[ -\tilde\omega^\mu P_\mu + \tilde\omega_Q^\alpha Q_\alpha + \bar{\tilde\omega}_{\bar{Q}\dot\alpha} \bar{Q}^{\dot\alpha} + \tilde\omega_D D + \tilde\omega_R R \right. \cr
 & &\left. \qquad\qquad\qquad\qquad\qquad\qquad- \tilde\omega_{\cal K}^\mu {\cal K}_\mu + \tilde\omega_S^\alpha S_\alpha + \bar{\tilde\omega}_{\bar{S}\dot\alpha} \bar{S}^{\dot\alpha}  +\frac{1}{2} \tilde\omega_M^{\mu\nu} M_{\mu\nu}\right].
\eea
which is just the Maurer-Cartan form with $v^\mu =0$. The individual tilded one-forms then take the form
\bea
&&\tilde\omega^\mu = dx^\nu A_\nu^{~\mu} e^{\varphi}\left( 1 -\frac{1}{m^2}\lambda_S^2 \bar\lambda_S^2 \right) +\frac{i}{m^2} \left(\lambda_S \sigma^\mu d\bar\lambda -d\lambda_S \sigma^\mu \bar\lambda_S \right) - \frac{2}{m^2} (\lambda_S \sigma^\mu \bar\lambda_S ) da  \cr
 & & \cr
 & & \qquad\qquad+\frac{2}{m^2}e^{\varphi}\left[e^{-ia}\lambda_S^2 (d\lambda \sigma^\mu \bar\lambda_{\bar{S}} ) +e^{+ia}\bar\lambda_{\bar{S}}^2 ( \lambda_S \sigma^\mu d\bar{\lambda}) \right] \cr
 & & \cr
&&\tilde\omega_D = d\varphi  +2i e^{\varphi-ia}( \lambda_S d\lambda) -2ie^{\varphi+ia}(\bar\lambda_{\bar{S}}d\bar\lambda) = \tilde\omega^m e^{-\varphi}\tilde\nabla_m \varphi   \cr
 & & \cr
&&\tilde\omega_R = da +3 e^{\varphi}\left[e^{-ia} (\lambda_S d\lambda) + e^{+ia}(\bar\lambda_{\bar{S}}d\bar\lambda)\right] +3 dx^\nu A_\nu^{~\mu}e^{\varphi}(\lambda_S \sigma_m \bar\lambda_{\bar{S}} )   \equiv dx^\mu \nabla_\mu a \cr
 & & \cr
&&\tilde\omega_Q^\alpha = e^{\varphi}\left(e^{-ia}d\lambda - dx^\mu A_\mu^{~\nu}(\bar\lambda_{\bar{S}} \bar\sigma_\nu )^\alpha \right)\cr
 & & \cr
&&\bar{\tilde\omega}_{\bar{Q}\dot\alpha} = e^{\varphi}\left(e^{+ia}d\bar\lambda + dx^\mu A_\mu^{~\nu}(\lambda_S \sigma_\nu)_{\dot\alpha}\right) \cr
 & & \cr
&&\tilde\omega_S^\alpha = d\lambda_S^\alpha +\frac{1}{2} d\varphi \lambda_S^\alpha -i da \lambda_S^\alpha -i dx^\mu A_\mu^{~\nu}e^{\varphi}(\lambda_S \sigma_\nu \bar\lambda_{\bar{S}} ) \lambda_S^\alpha \cr
 & & \cr
 & &~~~~+\frac{i}{6} (\lambda_S \sigma^\tau \bar\lambda_{\bar{S}} ) \epsilon_{\mu\nu\rho\tau} (\lambda_S \sigma^{\mu\nu})^\alpha e^{\varphi}dx^\lambda A_\lambda^{~\rho} -i e^{\varphi}[e^{-ia}\lambda_S d\lambda +2 e^{+ia} \bar\lambda_{\bar{S}}d\bar\lambda]\lambda_S^\alpha  \cr
 & & \cr
 & & ~~~~-\frac{i}{4}e^{\varphi-ia} \left( d\lambda \sigma^{\mu\nu} \lambda_S \right) (\lambda_S \sigma_{\mu\nu})^\alpha \cr
 & & \cr
&&\bar{\tilde\omega}_{\bar{S}\dot\alpha} =  d\bar\lambda_{\bar{S}\dot\alpha} +\frac{1}{2} d\varphi \bar\lambda_{\bar{S}\dot\alpha} +i da \bar\lambda_{\bar{S}\dot\alpha} +i dx^\mu A_\mu^{~\nu}e^{\varphi}(\lambda_S \sigma_\nu \bar\lambda_{\bar{S}} ) \bar\lambda_{\bar{S}\dot\alpha} \cr
 & & \cr
 & &~~~~+\frac{i}{6} (\lambda_S \sigma^\tau \bar\lambda_{\bar{S}} ) \epsilon_{\mu\nu\rho\tau} (\bar\lambda_{\bar{S}} \bar\sigma^{\mu\nu})_{\dot\alpha} e^{\varphi}dx^\lambda A_\lambda^{~\rho} +ie^{\varphi} [2e^{-ia} \lambda_S d\lambda +e^{+ia}\bar\lambda_{\bar{S}}d\bar\lambda ]\bar\lambda_{\bar{S}\dot\alpha}  \cr
 & & \cr
 & & ~~~~+\frac{i}{4} e^{\varphi+ia}\left( d\bar\lambda \bar\sigma^{\mu\nu} \bar\lambda_{\bar{S}} \right) (\bar\lambda_{\bar{S}} \bar\sigma_{\mu\nu})_{\dot\alpha} \cr
 & & \cr
&&\tilde\omega_{\cal K}^\mu = 2m^2 \left(  dx^\nu A_\nu^{~\mu} e^{\varphi} -\tilde\omega^\mu \right) \cr
 & & \cr
&&\tilde\omega_{M}^{\mu\nu} = e^{\varphi}\left[2\left( e^{-ia}d\lambda \sigma^{\mu\nu} \lambda_S - e^{+ia}d\bar\lambda \bar\sigma^{\mu\nu} \bar\lambda_{\bar{S}}\right) +2 \epsilon^{\mu\nu}_{\rho\tau} (\lambda_S \sigma^\tau \bar\lambda_{\bar{S}}) dx^{\lambda} A_\lambda^{~\rho}\right] ,\cr
 & & 
\label{omegatilde}
\eea
where the Akulov-Volkov vierbein is defined as
\be
A_\mu^{~\nu} =\left[ \delta_\mu^{~\nu} +i\left( \lambda \sigma^\nu \partial_\mu \bar\lambda -\partial_\mu \lambda \sigma^\nu \bar\lambda \right) \right].
\label{AV}
\ee

The vierbein $e_\mu^{~\nu}$ relates the coordinate differentials $dx^\mu$ to the covariant coordinate differentials $\omega^\nu$ according to
\be
\omega^\nu = dx^\mu e_\mu^{~\nu}.
\ee
Since the one-form $\tilde\omega^\mu$ can also act as a basis one-form, one can expand
\be
\omega^\mu = dx^\nu e_\nu^{~\mu}=\tilde\omega^\nu N_\nu^{~\mu} .
\label{N}
\ee
where using Eq. (\ref{omega}), $N_\nu^{~\mu}$ is extracted as 
\be
N_\nu^{~\mu} = \delta_\nu^{~\mu} + [\cos({\sqrt{v^2 /m^2}}) -1]P_{\parallel \nu}^{~~~\mu} (v) - e^{-\varphi}\tilde\nabla_\nu \varphi \frac{v^\mu}{m^2} \frac{\sin({\sqrt{v^2 /m^2}})}{\sqrt{v^2 /m^2}} .
\ee
It is also useful to define the vierbein $\tilde{e}_\mu^{~\nu}$ as
\be
\tilde\omega^\mu = dx^\nu \tilde{e}_\nu^{~\mu}.
\ee
so that
\be
e_\mu^{~\nu} =\tilde{e}_\mu^{~\rho} N_\rho^{~\nu} .
\label{etildeN}
\ee

Using the Akulov-Volkov vierbein, $A_\mu^{~\nu}$, the $\tilde\omega^\mu$ one-form can be expanded as 
\be
\tilde\omega^\mu = dx^\nu \tilde{e}_\nu^{~\mu} =  dx^\nu e^\varphi A_\nu^{~\rho}T_\rho^{~\mu} ,
\label{ehatT}
\ee
where $T_\nu^{~\mu}$ can be gleaned from Eq. ($\ref{omegatilde}$) as
\bea
T_\nu^{~\mu} &=& \delta_\nu^{~\mu} (1-\frac{1}{m^2} \lambda_S^2 \bar\lambda_{\bar{S}}^2 ) +\frac{i}{m^2} (\lambda_S \sigma^\mu {\cal D}_\nu\bar\lambda_{\bar{S}} -{\cal D}_\nu \lambda_S \sigma^\mu \bar\lambda_S )e^{-\varphi} - \frac{2}{m^2} (\lambda_S \sigma^\mu \bar\lambda_S ) {\cal D}_\nu a e^{-\varphi}\cr
 & & \cr
 & &   +\frac{2}{m^2}\left[\lambda_S^2 ({\cal D}_\nu  \lambda \sigma^\mu \bar\lambda_{\bar{S}} )e^{-ia} +\bar\lambda_{\bar{S}}^2 ( \lambda_S \sigma^\mu {\cal D}_\nu \bar\lambda )e^{ +ia} \right]  ~,
\eea
with ${\cal D}_m = {A}_m^{-1 \mu}\partial_\mu$.  Using Eqs. (\ref{etildeN}) and (\ref{ehatT}), the vierbein can be written in a product form as
\be
e_\mu^{~\nu} = e^\varphi{A}_\mu^{~\rho} T_\rho^{~\tau} N_\tau^{~\nu} .
\ee

Since $d^4x \det{e}$ is invariant and an invariant kinetic energy for the $R$-axion can be formed by contracting the covariant derivatives with the vierbein, an $SU(2,2|1)$ invariant action is constructed as
\bea
S &=& -\sigma \int d^4 x \det{e}\left(1 +  \nabla_\mu a e^{-1\mu}_{\rho}\eta^{\rho\tau}e^{-1\nu}_\tau \nabla_\nu a\right) \cr
 &=& -\sigma \int d^4 x e^{4\varphi} \det{A} \det{N}\det{T}\left(1 + e^{-2\varphi} \nabla_\mu a h^{\mu\nu}\nabla_\nu a\right) ~,
\eea
where
\bea
h^{\mu\nu} &=& N_\rho^{-1 \tau}T_\tau^{-1 \lambda} A_\lambda^{-1 \mu} \eta^{\rho\upsilon}  N_\upsilon^{-1 \kappa}T_\kappa^{-1 \sigma} A_\sigma^{-1 \nu}.
\eea
The determinant of $N$ can be then be explicitly evaluated giving
\be
\det{N} = \cos({\sqrt{v^2 /m^2}}) \left[ 1 +  e^{-\varphi}\tilde\nabla_\mu \varphi \frac{v^\mu}{m^2} \frac{\tan({\sqrt{v^2 /m^2}})}{\sqrt{v^2 /m^2}} \right] .
\ee

Since the action only depends on $v^\mu$ and not its derivatives, it is not an independent dynamical degree of freedom.  As such it can be eliminated\cite{IO} by setting the invariant one-form $\omega_D$ to zero. Solving this constraint equation then gives
\be
v_\mu \frac{\tan({\sqrt{v^2 /m^2}})}{\sqrt{v^2 /m^2}} = - e^{-\varphi}\tilde\nabla_\mu \varphi .
\ee
which, in turn, allows us to write
\bea
N_\nu^{~\mu} &=& \delta_\nu^{~\mu} + \left( \frac{1}{\sqrt{1+\frac{e^{-2\varphi}}{m^2}\tilde\nabla_\rho \varphi \eta^{\rho\tau} \tilde\nabla_\tau \varphi }}-1\right) P_{\parallel \nu}^{~~~\mu} (\tilde\nabla\varphi) \cr
 & & -\frac{1}{\sqrt{1+\frac{e^{-2\varphi}}{m^2}\tilde\nabla_\rho \varphi \eta^{\rho\tau} \tilde\nabla_\tau \varphi }} \frac{e^{-2\varphi}}{m^2}\tilde\nabla_\nu \varphi \tilde\nabla^\mu \varphi .
\eea
The superconformal Goldstinos, $\lambda_S$ and $\bar\lambda_{\bar{S}}$, are also not independent dynamical degrees of freedom but can be expressed in terms of derivatives of the Goldstinos $\lambda$ and $\bar\lambda$, and products of these with the Nambu-Goldstone bosons $\varphi$ and $a$.  The covariant constraint equation is obtained by setting to zero the fermionic one-forms $\omega_Q^\alpha =0$ and $\bar\omega_{\bar{Q}\dot\alpha}=0$.  Combining the various one-forms in (\ref{omega}) and ($\ref{omegatilde}$) the solution to these covariant constraints begin as
\bea
\lambda_S^\alpha &=& -\frac{1}{4} (\sigma^\mu \partial_\mu \bar\lambda)^\alpha +\cdots \cr
\bar\lambda_{\bar{S}\dot\alpha} &=& -\frac{1}{4} (\partial_\mu \lambda \sigma^\mu )_{\dot\alpha} +\cdots .
\eea

Substituting the above expression for $N_\nu^{~\mu}$, 
the invariant action then takes the form
\be
S = -\sigma \int d^4 x e^{4\varphi} \det{A}~\det{T} ~~\sqrt{1 + \frac{e^{-2\varphi}}{m^2}\tilde\nabla_\mu \varphi \eta^{\mu\nu} \tilde\nabla_\nu \varphi }~\left(1 + e^{-2\varphi} \nabla_\mu a h^{\mu\nu}\nabla_\nu a\right) .
\label{sc}
\ee
The action is an invariant synthesis of Akulov-Volkov and Nambu-Goto actions. Note that the pure dilatonic part of the action (obtained by setting the Goldstinos and $a$ to zero so that $A_\mu^{~\nu}=T_\mu^{~\nu}=\delta_\mu^\nu$) reproduces the previous action, Eq. (\ref{mba}), of the Minkowski space $M_4$ probe brane in $AdS_{5}$ without SUSY. As such, the dilaton describes the motion of the probe brane into the rest of the $AdS_5$ space. However, in this case, because of the spontaneous breakdown of the complete SUSY, there is no invariant that can be added to the action to cancel the vacuum energy as one was able to achieve in the non-supersymmetric Minkowski space probe brane case (c.f. Eq. (\ref{mba1})). It follows that the dilaton dynamics feels an $e^{4\varphi}$ potential. This contains a destabilizing term linear in $\varphi$ which drives the dilaton field $\varphi \rightarrow -\infty$. Since the dilaton describes the motion of the probe Minkowski $M_4$ brane into the remainder of $AdS_5$ space, it follows that the Minkowski space brane is driven to infinite boundary of $AdS_5$ space and the interior of the $AdS_5$ space cannot sustain the brane. Alternatively expressed, the spectrum cannot include both the Goldstino and the dilaton as Nambu-Goldstone modes. 

An alternate combination of broken generators $K^\mu = \frac{1}{m^2}(\hat{P}^\mu -2m^2{\cal K}^\mu )$ can also be defined. This leads to the 4-d superconformal algebra
\bea
&&[P^\mu, P^\nu]=0~;~[K^\mu, K^\nu] =0~;~[P^\mu, K^\nu]= 2i(\eta^{\mu\nu}D -M^{\mu\nu})\cr
&&[D, P^\mu]=iP^\mu ~;~[D, K^\mu]=-iK^\mu ~;~[D, M^{\mu\nu}]= 0\cr
&&[M^{\mu\nu}, P^\lambda]=i(\eta^{\mu\lambda}P^\nu-\eta^{\nu\lambda}P^\mu)~;~[M^{\mu\nu}, K^\lambda]=i(\eta^{\mu\lambda}K^\nu-\eta^{\nu\lambda}K^\mu) \cr
&&[M^{\mu\nu}, M^{\lambda\rho}]=i(\eta^{\mu\lambda}M^{\nu\rho}-\eta^{\mu\rho}M^{\nu\lambda}-\eta^{\nu\lambda}M^{\mu\rho}+\eta^{\nu\rho}M^{\mu\lambda})\cr
&&[R,P^\mu]= 0~;~[R, K^\mu]=0~;~[R, M^{\mu\nu}]= 0 \cr
&&[P^\mu, Q_\alpha ]= 0 ~;~[P^\mu, \bar{Q}_{\dot\alpha}] = 0 \cr
&&\{Q_\alpha, \bar{Q}_{\dot\alpha}\}= 2\sigma^\mu_{\alpha\dot\alpha}P_\mu ~;~\{S_\alpha,\bar{S}_{\dot\alpha}\} = 2\sigma^\mu_{\alpha\dot\alpha} K_\mu\cr
&&\{Q_\alpha, Q_\beta\} = 0 ~;~\{\bar{Q}_{\dot\alpha}, \bar{Q}_{\dot\beta}\} =0 \cr
&&[M^{\mu\nu}, Q_\alpha]= -\frac{1}{2}\sigma^{\mu\nu}~_\alpha~^\beta Q_\beta ~;~[M^{\mu\nu}, \bar{Q}_{\dot\alpha}]= -\frac{1}{2}\bar{\sigma}^{\mu\nu}~_{\dot\alpha \dot\beta} \bar{Q}^{\dot\beta } \cr
&&[R, Q_\alpha]= Q_\alpha ~;~[R, \bar{Q}_{\dot\alpha}] = -\bar{Q}_{\dot\alpha}\cr
&&[D, Q_\alpha]= \frac{i}{2}Q_\alpha ~;~[D, \bar{Q}_{\dot\alpha}]= \frac{i}{2}\bar{Q}_{\dot\alpha} \cr
&&[R,S_\alpha ] = -S_\alpha ~;~[R,\bar{S}_{\dot\alpha}] = \bar{S}_{\dot\alpha}\cr
&&[D,S_\alpha]= -\frac{i}{2}S_\alpha ~;~[D,\bar{S}_{\dot\alpha}]=-\frac{i}{2}\bar{S}_{\dot\alpha} \cr
&&[K^\mu, Q_\alpha]= i\sigma^\mu_{\alpha\dot\alpha} \bar{S}^{\dot\alpha} ~;~[K^\mu, \bar{Q}_{\dot\alpha}]= iS^\alpha \sigma^\mu_{\alpha\dot\alpha} \cr
&&[P^\mu, S_\alpha ]= i\sigma^\mu_{\alpha\dot\alpha}\bar{Q}^{\dot\alpha}~;~[P^\mu, \bar{S}_{\dot\alpha}]= iQ^\alpha \sigma^\mu_{\alpha\dot\alpha} \cr
&&[M^{\mu\nu}, S_\alpha]= -\frac{1}{2}\sigma^{\mu\nu}~_\alpha~^\beta S_\beta ~;~[M^{\mu\nu}, \bar{S}_{\dot\alpha}]= -\frac{1}{2}\bar{\sigma}^{\mu\nu}~_{\dot\alpha \dot\beta} \bar{S}^{\dot\beta } \cr
&&[K^\mu, S_\alpha]=0 ~;~[K^\mu,\bar{S}_{\dot\alpha}]=0 \cr
&&\{Q_\alpha,S_\beta \}= i (\sigma^{\mu\nu}_{\alpha\beta}M_{\mu\nu}+2i\epsilon_{\alpha\beta}D+3\epsilon_{\alpha\beta}R) \cr
&&\{\bar{Q}_{\dot\alpha}, \bar{S}_{\dot\beta} \}= -i (\bar\sigma^{\mu\nu}_{\dot\alpha\dot\beta}M_{\mu\nu}-2i\epsilon_{\dot\alpha\dot\beta}D + 3\epsilon_{\dot\alpha\dot\beta}R) \cr
&&\{Q_\alpha, \bar{S}_{\dot\alpha}\}= 0~;~\{S_\alpha, \bar{Q}_{\dot\alpha}\}= 0 ~;~\{S_\alpha, S_\beta \}=0 ~;~\{\bar{S}_{\dot\alpha}, \bar{S}_{\dot\beta}\} = 0 ~.
\eea

The spontaneously broken symmetries are
$R$, dilatations ($D$), special conformal ($K^\mu$), SUSY ($Q_\alpha~,~\bar{Q}_{\dot\alpha}$) and SUSY conformal ($S_\alpha~,~\bar{S}_{\dot\alpha}$). Since the generators $K^\mu$ and ${\cal K}^\mu$ differ only by unbroken translation generator $P^\mu$, the action ($\ref{sc}$) is invariant under superconformal transformations. The leading terms in a momentum expansion are just
\bea
&&S=-\sigma\int d^4x  \{e^{4\varphi}~det~A 
-\frac{1}{2} ~det~A~ e^{2\varphi}{\cal D}_\mu\varphi \eta^{\mu\nu}{\cal D}_\nu\varphi \cr
&&~~~~~~~~- \frac{1}{2}~det~A~e^{2\varphi} {\cal D}_\mu a \eta^{\mu\nu}{\cal D}_\nu a \} ~.
\eea
Once again the potential for the dilaton $\varphi$ is unstable and there is an incompatibility of simultaneous nonlinear realizations of SUSY and scale symmetry in Minkowski space\cite{CLtheorem}. Note that the origin of this unusual behavior is not simply a consequence of the introduction of a scale due the spontaneously broken SUSY. It has been shown that there is no incompatibility in securing simultaneous nonlinear realization of spontaneously broken scale and chiral symmetries\cite{B} where a scale is also introduced. In that case, the spectrum of the effective Lagrangian admits both a pion and a dilaton.\\
\\

\noindent This was supported in part by the U.S. Department of Energy under grant DE-FG02-91ER40681 (Task B). TEC thanks Muneto Nitta and Tonnis ter Veldhuis for numerous insightful discussions.

\end{document}